\documentclass[twocolumn]{aastex631}
\usepackage{amsmath}

\newcommand{\LarmorT}{\Omega_\mathrm{i}^{-1}}
\newcommand{\SkinI}{\lambda_\mathrm{si}}
\newcommand{\SkinE}{\lambda_\mathrm{se}}

\newcommand{\mpon}{\textcolor{black}}
\newcommand{\mponn}{\textcolor{black}}
\newcommand{\ab}{\textcolor{black}}
\newcommand{\kf}{\textcolor{black}}
\newcommand{\KF}{\textcolor{black}}
\newcommand{\kff}{\textcolor{black}}
\newcommand{\kfff}{\textcolor{black}}

\begin{document}

\title{Kinetic simulations of \kff{non-relativistic high-Mach-number} perpendicular shocks propagating in \kff{a} turbulent medium}

\correspondingauthor{Karol Fulat}
\email{karol.fulat@uni-potsdam.de}

\author[0000-0001-6002-6091]{Karol Fulat}
\affil{Institute of Physics and Astronomy, University of Potsdam, D-14476 Potsdam, Germany}

\author[0000-0002-5680-0766]{Artem Bohdan}
\affiliation{Max-Planck-Institut für Plasmaphysik, Boltzmannstr. 2, DE-85748 Garching, Germany}
\affiliation{Excellence Cluster ORIGINS, Boltzmannstr. 2, DE-85748 Garching, Germany}

\author[0000-0003-1699-5770]{Gabriel Torralba Paz}
\affil{Institute of Nuclear Physics PAN, Radzikowskiego 152,
31-342, Kraków, Poland}

\author[0000-0001-7861-1707]{Martin Pohl}
\affil{Institute of Physics and Astronomy, University of Potsdam, D-14476 Potsdam, Germany}
\affil{Deutsches Elektronen-Synchrotron DESY, Platanenallee 6, 15738 Zeuthen, Germany}

\begin{abstract}
\kff{Strong non-relativistic shocks are known to accelerate particles up to relativistic energies. However,} for Diffusive Shock Acceleration electrons must have a highly suprathermal energy, implying a need for very efficient pre-acceleration. Most published studies consider \KF{shocks propagating through homogeneous plasma}, which is an unrealistic assumption for astrophysical environments. Using 2D3V particle-in-cell simulations, we investigate electron acceleration and heating processes at \mponn{non-relativistic} high-Mach-number shocks \mponn{in electron-ion plasma} with a turbulent upstream medium. For this purpose slabs of plasma with \KF{compressive turbulence} are separately simulated and then inserted into shock simulations, which requires matching of the plasma slabs at the interface. Using a novel \kff{procedure}
of matching electromagnetic fields and currents, we perform simulations of perpendicular shocks setting different intensities of density fluctuations ($\lesssim10\%$) in the upstream. \kff{The new simulation technique provides a framework for studying shocks propagating in turbulent media.} We explore the impact of the fluctuations on electron heating, the dynamics of upstream electrons, and the driving of plasma instabilities. Our results indicate that while the presence of the turbulence enhances variations in the upstream magnetic field, their levels remain too low to influence significantly the behavior of electrons \mpon{at perpendicular shocks}.
\end{abstract}


\section{Introduction} \label{sec:intro}

Understanding how particles are accelerated at shock waves remains a fundamental challenge for astrophysics and space physics. The description provided by the basic process, diffusive shock acceleration \citep[DSA; e.g.,][]{Bell1978,Blandford1978,Drury1983,Blandford1987}, is not complete \mponn{and does not always describe the behavior} at interplanetary shocks \citep[see e.g.][]{Lario2003}. The influence of pre-existing turbulence upstream of a shock is an important \mpon{question} in the acceleration theory, yet is still poorly understood. It is known, that fluctuations enhance the trapping of particles near a shock, providing \mpon{favorable} conditions for effective acceleration \citep[see][]{Guo2021,Perri2022}. Turbulence is ubiquitous in astrophysical environments, thus it is essential to investigate its interplay with shock waves. \kff{In this paper we study collisionless shocks in the non-relativistic high-Mach-number regime, with the sonic and Alfv{\'e}nic Mach numbers $M_S, M_A \gtrsim 20$. \kfff{Such} conditions are observed mainly in supernova remnants \citep[SNRs,][]{Raymond2023}, but also in the bow shocks of Jupiter \citep{Slavin1985}, Saturn \citep{Sulaiman2016}, and Uranus \citep{Bagenal1987}, and occasionally at the Earth's bow shock \citep{Sundberg2017}.}

Since it is widely accepted that supernova remnants are plausible candidates for the production of the galactic cosmic rays, \kfff{we identify these sources as primary objects where our simulations may find application.} The \kff{fast} shocks, \kff{$v_\mathrm{sh} \approx 1,000 - 10,000$ km/s,} form during the interaction of supernova ejecta with the ambient medium, which is turbulent. \kff{The impact of the medium inhomogeneities on the evolution of SNRs is an important aspect for understanding the observed emission morphology. Previous numerical studies have demonstrated that magnetohydrodynamic (MHD) turbulence modify the dynamical properties of these systems \citep{Balsara2001,Zhang2019,Peng2020,Bao2021} and amplify the magnetic field \citep{Giacalone2007,Guo2012a}.}



Observations of \kf{radio synchrotron emission \citep[e.g.,][]{Shklovsky1954,Dubner2015} and} non-thermal X-ray emission \citep[e.g.,][]{Koyama1995,Vink2012}, as well as in the $\gamma$-ray waveband \citep[e.g.,][]{Aharonian2007}, indicate the presence of accelerated electrons in SNRs. The straightforward explanation for their energy gain is participation in DSA. However, thermal electrons do not satisfy the main requirement of the mechanism: their Larmor radii are much smaller than the shock width. For this reason they must undergo some pre-acceleration mechanism before being injected into DSA, which is known as the electron injection problem. Possible mechanisms and the underlying micro-physics of electron acceleration have been extensively investigated \citep[see][for a review]{Amano2022,Bohdan2023}. \kff{To interpret the broadband emission from supernova remnants, MHD simulations coupled with a prescription of particle acceleration have been utilized \citep[see e.g.,][for a review]{Orlando2021}. Such studies, e.g., \citet[][]{Ferrand2010,Orlando2012,Ferrand2014,Pavlovic2017,Pavlovic2018,Brose2021}, \kfff{usually employing Blasi's semi-analytical model of non-linear DSA} \citep{Blasi2002,Blasi2004,Blasi2005}, which \kff{treats} the electron injection efficiency as a free parameter.}

Previous studies of electron acceleration at the \kff{non-relativistic} shocks assumed a homogeneous upstream medium, hence all the turbulence in this region was driven only by shock-reflected particles. It is important to investigate the effect of pre-existing \kff{small-scale} turbulence in \kff{the high-Mach-number} regime and compare the previous simulations with new ones, where the upstream medium initially carries fluctuations. Shock waves propagate there through a turbulent medium, which apart from magnetic fluctuations carries density inhomogeneities. While the magnitude of the density fluctuations at SNRs remains unknown, measurements in the heliosphere \citep{Carbone2021} \KF{and in the local interstellar medium} \citep{Lee2020,Ocker2021,Fraternale2022} show amplitudes of $\delta n_e/n_e \lesssim 10\%$ on minute timescales in the spacecraft frame, which roughly corresponds to the kinetic scales of electrons.

The interaction of a shock with pre-existing inhomogeneities has been intensively studied in the magnetohydrodynamics (MHD) regime \citep[e.g.][]{Inoue2013, Mizuno2014, Hu2022}. The results show that turbulence is able to distort the shock structure: for instance, it changes the compression ratio and \mpon{induces rippling of} the shock surface. Furthermore, it amplifies the magnetic field via turbulent dynamo. Recent studies on ion scales via hybrid kinetic simulations demonstrate the strong influence of upstream turbulence on particle transport at shocks, as well as efficient proton acceleration \citep{Trotta2021, Nakanotani2022}. In the hybrid simulations electrons are treated as fluid. \citet{Guo2010,Guo2012b} \mpon{investigated their acceleration processes  with test-particle simulations and} found, that pre-existing magnetic turbulence significantly improves the electron-acceleration efficiency.

To fully examine the physics of shock waves propagating in turbulent plasmas, particularly on electron kinetic scales, particle-in-cell (PIC) simulations are needed. This method describes collisionless plasma from first principles, hence it is widely used in studies of shock waves in astrophysics \citep[e.g.][]{Pohl2020}. So far, first attempts to investigate the effects of a nonhomogeneous medium on plasma shocks via kinetic simulations have been made for relativistic shocks propagating in pair plasmas. Static large-scale variations of the upstream density significantly modify the conditions of the downstream plasma \citep{Tomita2019} and corrugate the shock front, which leads to efficient particle acceleration \citep{Demidem2023}. With a more realistic setup, in which the upstream plasma contains turbulence instead of \mpon{static density structures}, \citet{Bresci2023} confirmed an increased acceleration efficiency \mponn{at relativistic shocks} in the case of strong \mpon{electromagnetic} fluctuations. However, the statistical properties of the turbulence varied throughout the simulation. 

In this study, we investigate the effect of pre-existing turbulence \kff{at kinetic scales} on electron acceleration at non-relativistic \kff{high-Mach-number} shocks in electron-ion plasma using a novel technique. In our simulations the upstream medium is fully filled with turbulence, which has stable \mpon{statistical properties, meaning the intensity and spectrum of the fluctuations do not significantly} vary during the evolution of the system.

\mpon{Besides the} Mach numbers the main factor determining the physics of \kff{non-relativistic high-Mach-number} shocks is the inclination of the large-scale magnetic field. \mpon{Using} the obliquity angle $\theta_{\mathrm{Bn}}$ between the magnetic field and the shock normal, shocks can be divided into perpendicular ($\theta_{\mathrm{Bn}}=90^\circ$), oblique ($\theta_{\mathrm{Bn}}\approx50^\circ-75^\circ$), and quasi-parallel ($\theta_{\mathrm{Bn}}=0^\circ$). In this work, we focus on well-studied perpendicular shocks \citep{Matsumoto2012,Matsumoto2013,Matsumoto2015,Wieland2016,Bohdan2017,Bohdan2019a,Bohdan2019b,Bohdan2020b,Bohdan2020a}. The perpendicular direction of the magnetic field \mpon{confines the shock front to a thickness} of roughly one ion gyroradius, \mpon{and} the turbulence driven by reflected particles is \mpon{negligible}. The global structure of the perpendicular shocks is highly influenced by the electromagnetic Weibel instability \citep{Fried1959}, growing from the interaction between incoming and reflected, \mpon{in fact gyrating,} ions \citep{Kato2010}. \ab{The Weibel} instability \ab{strongly} amplifies the magnetic field. The model of this process provided by \citet{Bohdan2021} is consistent with in-situ measurements at Saturn's bow shock. The nonlinear evolution of the Weibel modes creates current sheets, which decay at the shock ramp through magnetic reconnection \citep{Matsumoto2015,Bohdan2020a}. The turbulence introduced by this process creates favorable conditions for electron acceleration. \mpon{At the leading edge of the shock foot, the Buneman instability is driven by} the interaction between cold incoming electrons and reflected ions \citep{Buneman1958}, \mpon{and electrons can} undergo shock-surfing acceleration \citep[SSA;][]{Shimada2000,Hoshino2002}. However, simulations show that most of the non-thermal particles found in the downstream region gained their energy via stochastic Fermi acceleration \citep[SFA;][]{Bohdan2017,Bohdan2019b}. \kff{In this process, the maximum energy that electrons can achieve is 10-100 times higher than the thermal energy of the downstream electrons. This is  insufficient for injection into DSA. Studies involving hybrid simulations \citep{Caprioli2014} and PIC-MHD \citep{vanMarle2022} have also indicated that perpendicular shocks are not efficient particle accelerators, except \kff{when the} magnetic field is low, \kff{or} seed cosmic rays \citep{Caprioli2018} or neutral particles \citep{Ohira2016} are present. Magnetic reconnection, SSA and SFA,} together with the shock potential and adiabatic compression, heat up electrons.

The paper begins \KF{with a} description \KF{of the simulations and the used setup} (Section \ref{sec:methods}). It \KF{follows with a} presentation of the results in Section \ref{sec:results}. Finally, a summary and discussion \KF{can be found} in Section \ref{sec:summary}.

\section{Methods} \label{sec:methods}
In the study presented here, we use the \mpon{MPI-parallelized code THATMPI \citep{Niemiec2008,Bohdan2017} that tracks 2-spatial and all 3 velocity components of individual particles in a plasma (2D3V configuration). This approximation} saves computational resources in comparison with the full 3D simulations, but reasonably accurately reproduces the physics of shock waves \citep[see, e.g.,][]{Pohl2020}. 

\mpon{We establish the shock by reflecting} an incoming plasma beam off a conducting wall. The interaction of the incoming and the reflected beam creates the shock. Since the simulation \mpon{is performed in} the downstream frame, the upstream plasma which flows toward the shock has to be continuously replenished. Homogeneous plasma can be added in a thin layer. However if that plasma is meant to carry pre-existing fluctuations, the turbulence has to be established separately and then injected into the shock simulation. In the following paragraphs, the method of turbulence generation is described, as well as its injection procedure into a shock simulation. Details of both are presented in the Appendix.

\subsection{Turbulence driving}
In the standard models of MHD turbulence the energy cascades from large scales to small scales \citep{Sridhar1994,Goldreich1995}. The 
separation of these scales makes it computationally infeasible to follow the self-consistent evolution of plasma turbulence down to kinetic scales. The Langevin antenna, a method of driving turbulence presented in \citet{Tenbarge2014}, allows mimicking the MHD cascade at scales resolved by PIC simulations \citep{Zhdankin2017,Groselj2019}. Despite this advantage, the method does not provide a precise control of plasma conditions (they vary in time). Furthermore, turbulence has to be driven continuously in some part of a simulation box, which enforces the use of elongated boxes for simulations in the downstream frame \citep[see][]{Bresci2023}. For \mpon{these reasons, we choose the so-called decaying turbulence and inject a set of modes at the beginning of a simulation that evolve into} turbulent structures. The following paragraph shows why the decaying setup is more suitable than the Langevin antenna for our case. 

Our study focuses on compressive turbulence.  We pre-fabricate slabs of \mpon{turbulent plasma in dedicated} simulations with a square simulation box and periodic boundaries. Their size corresponds to the width of the shock simulation into which they are to be injected. \mpon{In the pre-fabrication stage} density fluctuations are obtained by superposition of wave-like disturbances of the local bulk velocity \citep{Giacalone1994,Groselj2019}; their exact form is included in Appendix A. Fluctuations of the magnetic field evolve self-consistently, but are weaker than those in density. Even though the decaying turbulence cannot achieve a steady state, its statistical properties vary \mpon{very slowly during} the shock simulation. \ab{After a short period of rapid evolution} the magnitude of the density fluctuations is stable for \mpon{at least two ion Larmor times}, which means that they are sufficiently long-lived to be inserted into the shock simulation.

On kinetic scales, damping transfers wave energy into plasma heat. \mpon{Thus, we have to launch the pre-fabrication of plasma slabs with a low initial temperature and make sure that the temperature at the time of insertion into the shock simulation corresponds to the desired high Mach number.} The \mpon{initial heating constrains the achievable amplitude of density fluctuations to $\delta n/n \approx 10\%$}, where $\delta n$ is the root-mean-square of the particle density fluctuations (on the scale of a quarter of ion skin length), and $n$ is the mean density.

\subsection{Turbulence injection}
\mpon{For efficiency and for maintaining a similar evolution state of the turbulent upstream plasma we regularly add the pre-fabricated plasma slabs to our shock simulation.} Each new slab must be matched to the adjacent plasma (at the end of the shock simulation box) to prevent \mpon{artefacts and transients}. Here we briefly present our novel matching technique; a more detailed description can be found in Appendix B.

The matching is done for each computational cell in a selected region called the interpolation zone. We interpolate the magnetic and the electric fields, and then we \mpon{re-initiate the particles to reproduce the first and second} moments of the interpolated distribution function. \mpon{The zeroth moment of the interpolated distribution must correspond to an integer number of simulated particles. This leads to noise in the charge density that we balance by additional electric field to satisfy Gauss' law. A} nonzero divergence of the magnetic field is cleaned by the projection method \citep[see][]{Zhang2016}. After these two corrections the plasma self-consistently evolves without any numerical transients.

\begin{table}[ht!]
    \begin{center}
        \caption{The density fluctuation \mpon{amplitudes} and the measured shock properties \mpon{in our simulations.}}
        \begin{tabular}{l|c|c|c}
            \hline
            \hline
            Run & H & T1 & T2 \\
            \hline
            $\delta n/n \ [\%]$ & - & 3.5 & 10 \\
            $v_\mathrm{sh}/c \ [\times10^{-3}]$ & $263\pm2$ & $265\pm3$ & $263\pm3$ \\
            $k_BT_e/m_e c^2 \ [\times10^{-3}]$ & $168\pm4$ & $165\pm6$ & $170\pm6$ \\
            \hline
        \end{tabular}
        \label{tab:sim_params}
    \end{center}
    \tablecomments{All simulations have the same sonic and Alfv{\'e}nic Mach number, $M_S\approx36$ and $M_A \approx 32$, likewise the plasma beta, $\beta_p \approx 1$. The width of the simulation box for each case is roughly $L_y \approx 6\SkinI$. The shock velocities $v_\mathrm{sh}$ are measured in the upstream frame, and the electron temperatures $T_e$ are measured in the downstream region.}
\end{table}

\subsection{Simulation setup} \label{sec:setup}
To examine the influence of compressive pre-existing turbulence on perpendicular shocks we performed three simulations, one with homogeneous upstream medium (run H), and two with different \mpon{amplitudes} of density fluctuations, $\delta n/n =3.5\%$ (run T1) and $\delta n/n =10\%$ (run T2). \mpon{For easy comparison} with previous kinetic simulations of SNR shocks we used the same plasma parameters as for run B2 of \citet{Bohdan2019a}, except for a smaller width of the simulation box, $L_y \approx 6 \SkinI$ instead of $L_y = 24 \SkinI$. This \mpon{choice is computationally cheaper, but some features like} shock front corrugations may not be resolved. Nevertheless, the size is large enough to probe the influence of the turbulence on the electron scale physics, which is the main aim of this work. Table \ref{tab:sim_params} shows the turbulence amplitude and other important plasma parameters for each simulation. \mpon{The amplitude of the density fluctuations, $\delta n$, is calculated as the RMS of the particle density measured in small tiles that are a quarter of ion skin length in size. }

We initialize the upstream plasma with twenty particles per cell for both ions and electrons, \KF{$n_0=n_i=n_e=20$}\mponn{, where the subscript ``$i$'' represents ions and ``$e$'' electrons}.
The plasma flows toward the reflecting wall with velocity $\mathbf{v_0}=v_0\hat{x}$, where $v_0=-0.2c$. The large scale magnetic field is perpendicular to the shock normal, $\theta_0=90^\circ$, and has the in-plane configuration, $\mathbf{B}_0=b_0 \mathbf{\hat{y}}$. This magnetic field orientation leads to particle gyration in the $xz$ plane, \kf{giving particles} three degrees of freedom and an adiabatic index $\Gamma=5/3$.
\mpon{The expected shock speed in the upstream frame is $v_\mathrm{sh} \simeq 0.264c$.}

\mpon{In the shock simulation the particle} species are in quasi-equilibrium \kf{$T_e \approx T_i \approx 1.6\cdot10^{-3}m_ec^2/k_B\approx 10^7 \text{ K}$}. \mpon{For such a non-relativistic thermal distribution of particles the thermal speed} of electrons is defined as $v_{th,e} = \sqrt{2k_BT_e/m_e}\simeq 0.057c$. \mpon{On account of the ion-to-electron mass ratio $m_i/m_e=100$, the} thermal speed of ions is ten times smaller. The plasma beta, defined as the ratio of the thermal pressure to the magnetic pressure, \mpon{is $\beta_p \approx 1$ for all runs. The sonic Mach number then follows as $M_S=v_\mathrm{sh}/c_S\approx 36$ for all runs, where $c_S=\sqrt{\Gamma k_B(T_e+T_i)/m_i}\approx 0.0074c$. Likewise, the Alfv{\'e}n speed $v_A=B_0c/\sqrt{n(m_e+m_i)}=0.00829c$ leads to an  Alfv{\'e}nic Mach number $M_A \approx 32$.} \kf{The expected compression ratio at the shock is $r \approx 3.97$.}  

The time step, \mpon{$\delta t=1/40 \omega_{pe}^{-1}$, scales with the electron plasma frequency, $\omega_{pe}=\sqrt{q_e^2n_e/(\epsilon_0m_e)}$. We evolve the system for ten ion cyclotron times}, $\LarmorT=m_i/(q_i b_0) \approx 48,000 \delta t$. The electron \mpon{skin depth is $\SkinE=c/\omega_{pe}=20 \Delta$, where $\Delta$ represents the cell size;} for ions it is higher by a factor $\sqrt{m_i/m_e}$, which gives $\SkinI=200 \Delta$.

\kff{The chosen values for our simulation parameters are a compromise between the capabilities of modern supercomputers and the accurate modelling of \kff{the shocks}. By using a shock speed that is higher than typically observed in \kff{Galactic} sources, \kff{but still non-relativistic,} we are able to extend the duration of our simulations. In addition, these parameter choices allow for a direct comparison with previous studies of high-Mach-number shocks.}

\section{Simulation results} \label{sec:results}
Here, \KF{we compare simulations} with different intensities of upstream turbulence, see Table \ref{tab:sim_params}. Firstly, we discuss the global structure of a shock wave propagating in pre-existing turbulence, \mpon{including} the shock-reformation period, the shock speed, and the magnetic-field amplification. Then we focus on the shock foot, where we investigate how the density fluctuations influence the Buneman instability. Afterwards, we examine the Weibel instability together with magnetic reconnection. Further, we discuss the vorticity in the context of magnetic-field amplification via the turbulent dynamo and how the orientation of the magnetic field is modified by pre-existing fluctuations. Finally, we present the electron energy spectra along with the temperature of the particles.

\begin{figure*}[ht!]
\includegraphics[width=\textwidth]{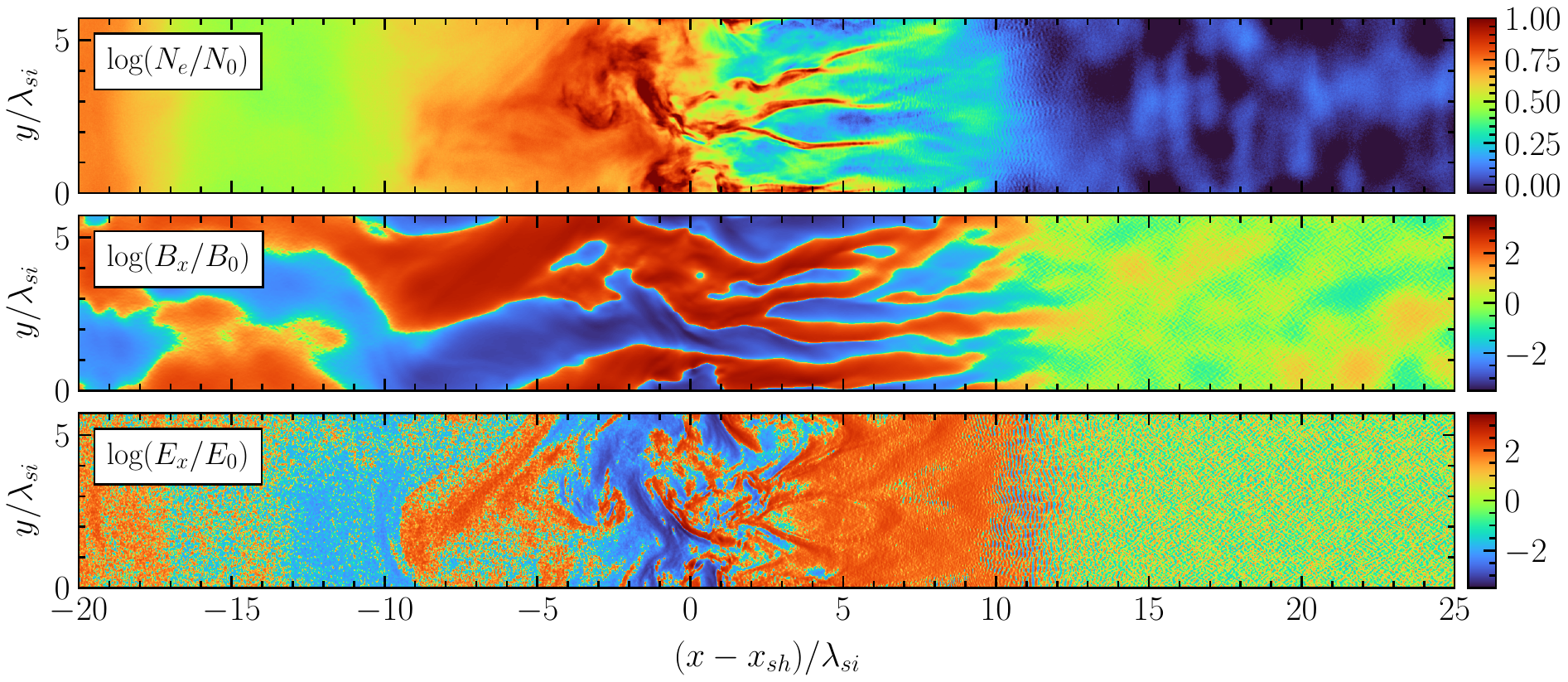}
\caption{Maps of the electron number density (\textit{top} panel) and the $x$-components of the magnetic (\textit{middle} panel) and the electric fields (\textit{bottom} panel) at a shock with pre-existing upstream density fluctuations (run T2). The scaling of the fields is logarithmic and sign-preserving, e.g. for $B_x/B_0$ it is $\text{sgn}(B_x)\cdot[2+\log \{\text{max}(10^{-2},|B_x|/B_0)\}]$.
\label{fig:maps2D}}
\end{figure*}

\begin{figure}[ht!]
\includegraphics[width=\columnwidth]{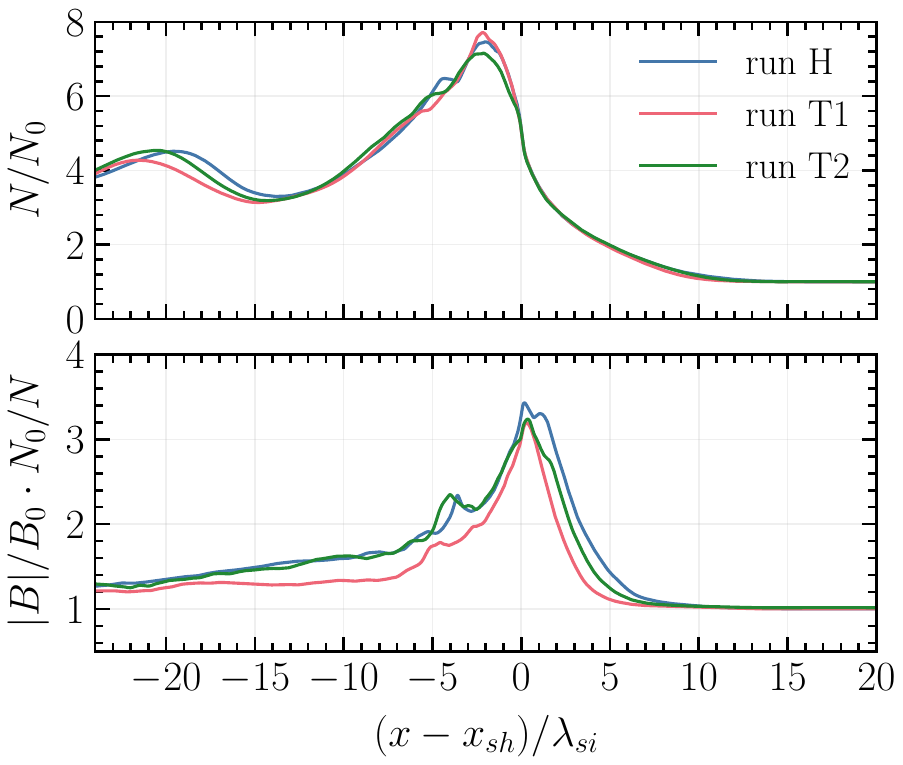}
\caption{The ion number density profile averaged over the $y$-direction \kf{(\textit{top} panel)}, as well as the ratio of the magnetic-field strength to \kf{the density} \kf{(\textit{bottom} panel)}. 
\label{fig:profiles}}
\end{figure}

\begin{figure}[ht!]
\includegraphics[width=\columnwidth]{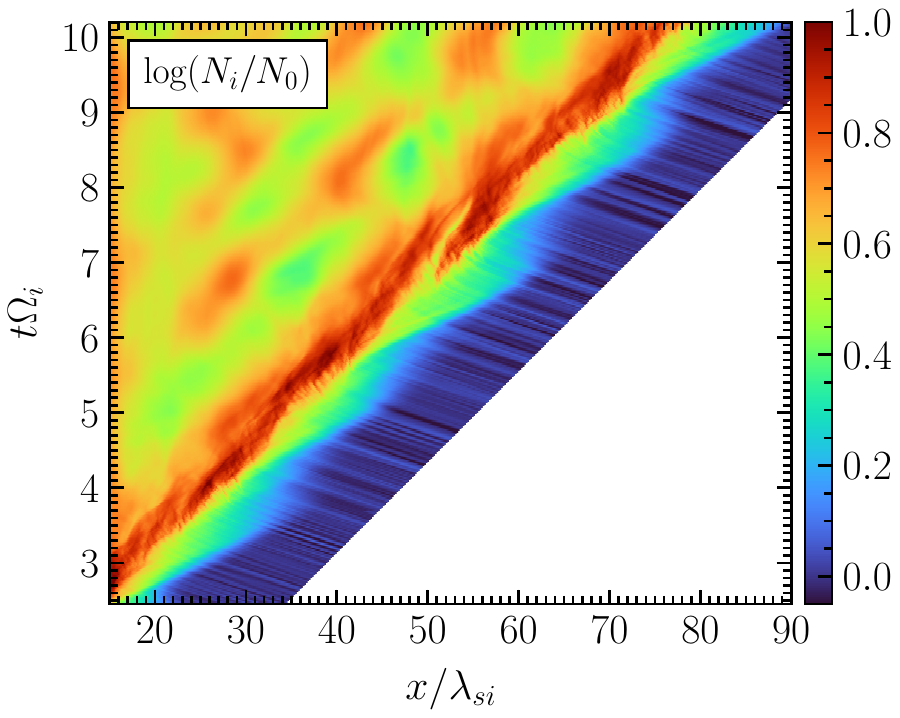}
\caption{The evolution of the ion number density averaged over the $y$-direction for run T2. \kf{Dark blue stripes represent upstream density fluctuations, which are advected to the shock front. For visualization purposes, the \mpon{presentation} is truncated $20\SkinI$ ahead of the shock.
}
\label{fig:dens}}
\end{figure}

\subsection{Global shock structure}\label{sec:global_structure}
The structure of high-Mach-number supercritical shocks is \mpon{determined by the reflection of upstream ions at the shock front, which for perpendicular shocks simply is a gyration in the shock-compressed magnetic field and leads to the formation of} the shock foot, the shock ramp, and the overshoot-undershoot structure downstream of the shock. \kf{Figure~\ref{fig:maps2D} shows the structure of a shock propagating in turbulent upstream medium (run T2), at time $t\approx10\LarmorT$. The upstream plasma at $x-x_{sh} \gtrsim 12\SkinI$ carries \mpon{density fluctuations and} significantly weaker fluctuations of the magnetic field. The shock foot spans from $10\SkinI$ to $12\SkinI$ ahead of the shock front. It contains \mpon{electrostatic} waves excited by the electrostatic Buneman instability, apparent in the $E_x$ map. The foot region is followed by the ramp, which extends to the overshoot at $x-x_{sh} \approx -1\SkinI$. This region is dominated by the Weibel instability, which forms filamentary structures in the \mpon{density and in the magnetic field. The overshoot-undershoot pattern about $10\SkinI$ behind the shock front marks the transition} to the downstream region.}

\mpon{The top panel of Figure~\ref{fig:profiles} compares the ion-density profiles of all runs. Regardless of the intensity of the upstream density fluctuations, neither the characteristic shock structure nor the compression ratio are affected by the upstream turbulence. The compression is always} consistent with the expected ratio of $r=3.97$.

Figure \ref{fig:dens} shows the evolution of the density profile averaged over the $y$-direction for a simulation with \kf{turbulent upstream plasma (run T2). For visualization purposes, the \mpon{false-color representation} is truncated $20\SkinI$ ahead of the shock. The pre-existing turbulent structures are advected to the shock front, thus they appear as dark blue stripes on the plot.} The \mpon{cyclic shock reformation seen as} modulations of the shock front is a common feature of high-Mach-number shocks \citep[see e.g.,][]{Wieland2016}. \mpon{They are caused by the non-stationary reflection of ions: the shock} re-develops with \mpon{an average period of $1.64 \LarmorT$ for all the runs, similar to the $1.50 \LarmorT$ reported in \cite{Wieland2016} and the $1.55 \LarmorT$ found} in \cite{Bohdan2017}. \mpon{A corresponding quasi-periodic modulation is seen in} the shock speed and in the plasma density. \mpon{Reflected particles reach at most $12 \SkinI$ ahead of the shock front, and the extent of the shock foot is the same for homogeneous and for turbulent upstream plasma}. We do not observe any \mpon{of the shock distortions that were seen in} previous \kf{hybrid and kinetic simulations \citep[e.g.][]{Nakanotani2022,Demidem2023}.} \mpon{There the density structures were much} larger than the shock width, \kf{whereas here the largest density clumps are with roughly $3\SkinI$ considerably smaller than the width of the shock transition, about $15\SkinI$.}

In Table \ref{tab:sim_params} we present the shock speed in the upstream frame, calculated over three fully developed reformation cycles. Within the uncertainties, the shock speed does not vary with the upstream density fluctuation level \mpon{and is} consistent with the \mpon{expected} value of $0.263c$.

The bottom panel in Figure \ref{fig:profiles} presents the magnetic-field strength \mpon{normalized by the particle density to remove the effect of compression.} The \mpon{amplification of the magnetic field reaches a factor} 3.3 at the overshoot regardless of the intensity of the pre-existing fluctuations, \mpon{which is consistent with the values} reported in \citet{Bohdan2021}. The findings suggest that upstream density fluctuations with \mpon{amplitudes} below 10\% \mpon{do not modify the amplification of} the magnetic field.

\subsection{Buneman instability}\label{sec:buneman}

\begin{figure}[ht!]
\includegraphics[width=\columnwidth]{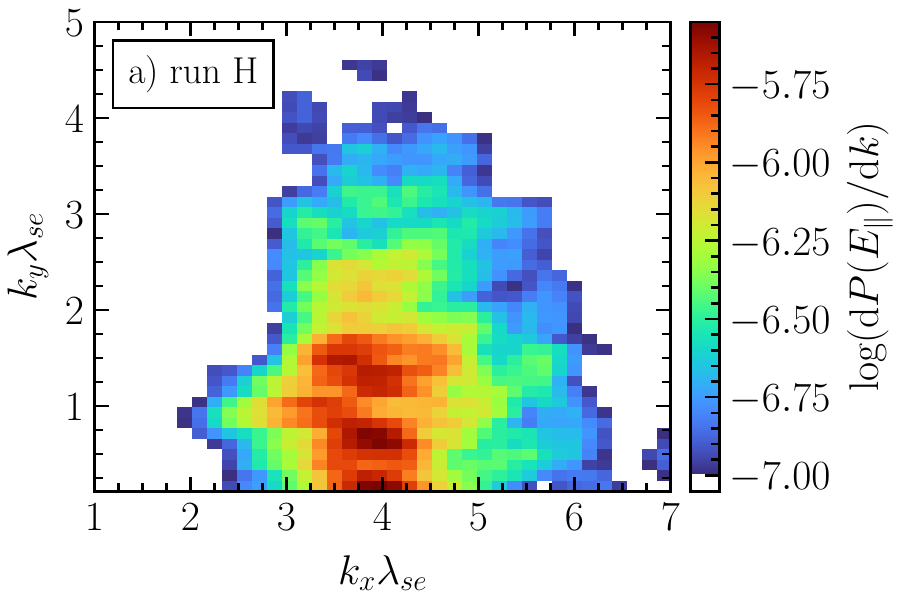}
\includegraphics[width=\columnwidth]{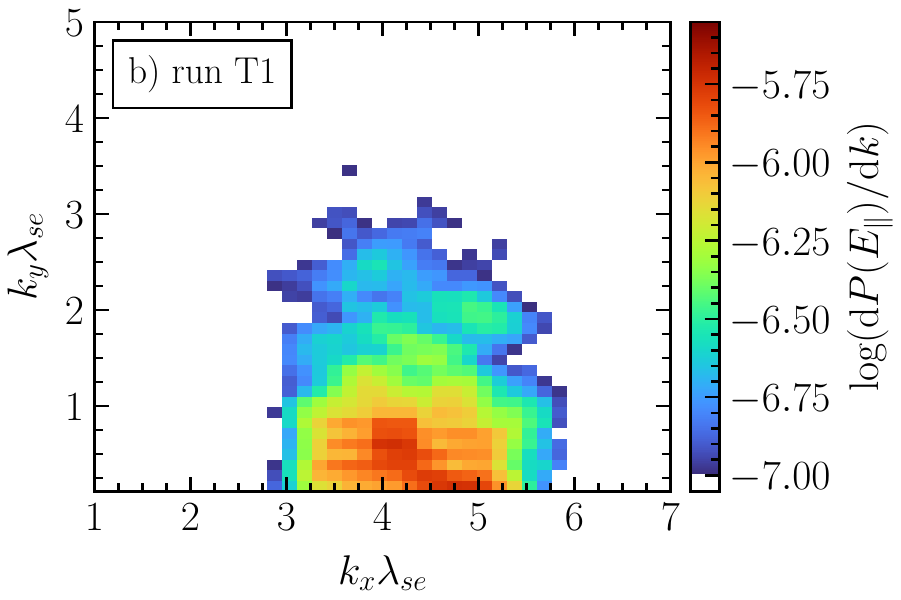}
\includegraphics[width=\columnwidth]{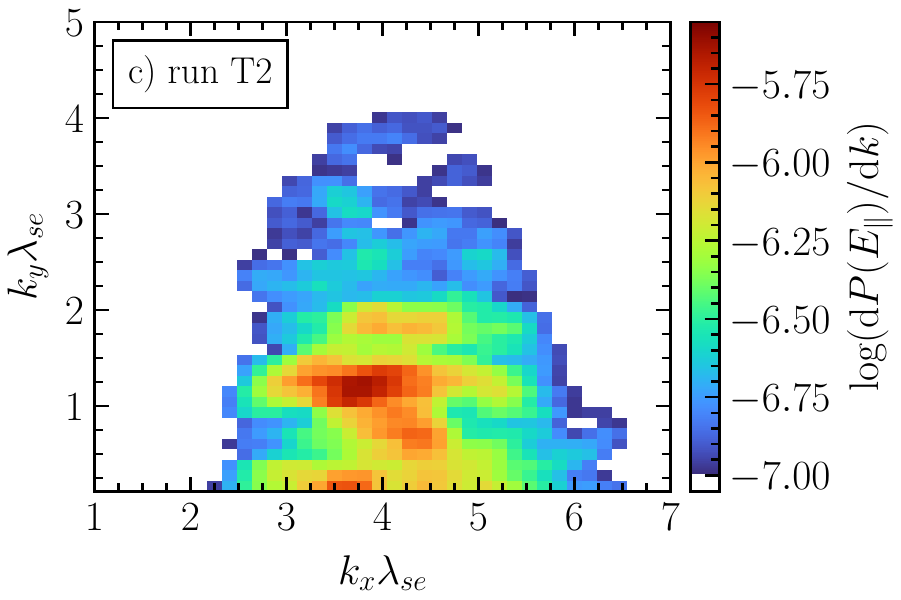}
\caption{The power spectrum of the electric field parallel to the wave vector, $\mathbf{E} \parallel \mathbf{k}$, for all runs, in the region where Buneman waves are the strongest during the latest reformation cycle. \kf{The power spectrum is shown in units of $(\omega_{pe}m_ec/e)^2$.} 
\label{fig:buneman}}
\end{figure}

Interaction of the shock-reflected ions with the incoming electrons causes the growth of electrostatic \ab{Buneman waves} in the shock foot. They play an important role in electron \mpon{heating as trapped particles} undergo shock surfing acceleration. We investigate how these modes are modified by the pre-existing upstream density fluctuations. We performed a Fourier analysis in the region where the electrostatic field has the largest amplitude during the latest reformation cycle.

Figure \ref{fig:buneman} \mpon{compares} the power spectrum of the electric field parallel to the wave vector, $E_{\parallel} \equiv \mathbf{E} \parallel \mathbf{k}$, for runs with homogeneous (run H, top) and turbulent \mpon{upstream plasma} (run T2, bottom). The evolution of the Buneman waves is non-stationary due to the shock reformation, \mpon{and} to compare the activity of the modes in different simulations, we chose a region where they have the largest magnitude during the last reformation cycle. Before calculation of the power spectrum, we remove the large-scale gradients from the electric field maps. The spectra have a \mpon{well-defined peak in $k_x$ and are broader} in the $k_y$-direction, which is in agreement with the multidimensional analysis of the Buneman instability \KF{\citep[see e.g.,][]{Amano2009}}. There are no significant differences between the spectra \mpon{and the electrostatic energy density between our runs, when we average over reformation cycles}: \kff{$8.0\cdot10^{-4}n_em_ec^2$} for run~H, \kff{$5.5\cdot10^{-4}n_em_ec^2$} for run~T1 and \kff{$6.5\cdot10^{-4}n_em_ec^2$} for run~T2.

\KF{The relative flow of ions and electrons can be characterized by the velocity difference of the respective beams $\bf{\Delta}\textbf{v}$. Here, we consider the cold beam of incoming electrons moving along the $x$-axis. The shock-reflected ions gyrate in the $xz$ plane, since in our simulations the perpendicular magnetic field has the in-plane configuration. \mponn{From the linear theory of the Buneman instability, the wave spectrum is dominated by modes parallel to $\bf{\Delta}\textbf{v}$:}
\begin{equation}
    k\SkinE \approx \frac{c}{\Delta v}
\label{eq:Buneman}
\end{equation}
\KF{where $\Delta v$ is the magnitude of the velocity \mpon{of reflected ions relative to the} incoming electrons \citep{Lampe1974}.} In 2D3V simulations, however, the $z$-component of the streaming motion \mponn{does not} drive waves, because the wavevectors must lie in the simulation plane. As a result, the \mponn{relevant} streaming motion of ions is primarily in the $x$-direction, yet the particles are warm (they gain energy at the shock): the $y$-component of $\bf{\Delta}\textbf{v}$ is not negligible. Therefore, in our simulations, we expect the waves to be slightly oblique with respect to the $x$-axis. Moreover, Equation \ref{eq:Buneman} \mponn{applies to} the fastest growing mode, \mponn{meaning that signal over a range} of wavenumbers perpendicular to the relative flow is present in the spectrum.}

\KF{In this paragraph we investigate whether the wave power spectra from Figure \ref{fig:buneman} are consistent with the  predictions of linear theory. To calculate the expected wavenumbers from Equation \ref{eq:Buneman} we estimate $\Delta v$ directly from the distribution functions of both plasma species. The relative speed for run H is approximately equal to $\Delta v\approx0.25c$, thus we expect the strongest modes with $k\SkinE\approx4.0$. The position of the peak in the $E_{\parallel}$ power spectrum shown in Figure \ref{fig:buneman}a is $\textbf{k}\SkinE=(3.8,1.5)$, so the magnitude $|\textbf{k}|\SkinE\approx4.1$ matches the wavenumber expected from Equation \ref{eq:Buneman}. The relative speed computed for run T1 approximately equals $\Delta v\approx0.24c$, which leads to the following wavenumber: $k\SkinE\approx4.2$. The strongest signal in Figure \ref{fig:buneman}b is found at $\textbf{k}\SkinE=(4.2,0.8)$, hence the magnitude is roughly equal to $|\textbf{k}|\SkinE\approx4.3$. This again corresponds to the predicted value. For run T2 the relative speed is roughly $\Delta v\approx0.24c$, which gives $k\SkinE\approx4.2$. Figure \ref{fig:buneman}c shows the peak at $\textbf{k}\SkinE=(3.8,1.2)$, which leads to $|\textbf{k}|\SkinE\approx4.0$. In this case, the linear theory predictions are also roughly in agreement with the power spectrum obtained from the simulation.}

\subsection{Weibel instability}\label{sec:weibel}

\begin{figure}[t!]
\includegraphics[width=\columnwidth]{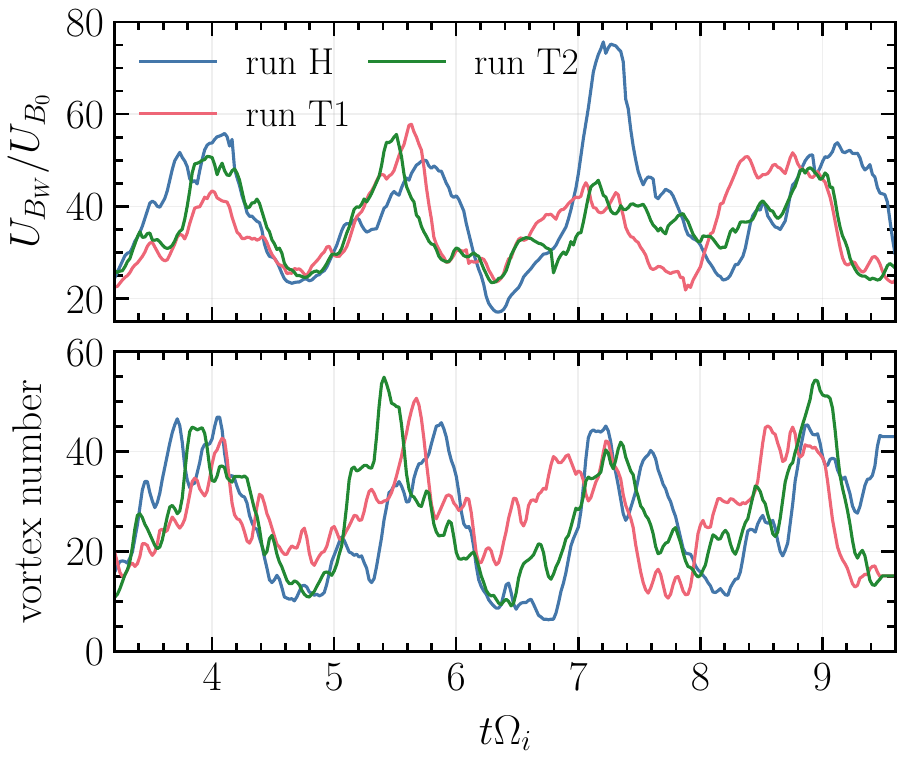}
\caption{\KF{The evolution of: the energy density of the Weibel-generated magnetic field (\textit{top} panel) and number of the magnetic vortices (\textit{bottom} panel), for all simulations. The energy density is normalized by the factor $U_{B_0}=|B_0|^2/(2\mu_0)$. The vortex number corresponds to the number of local maxima in $z$-component of the magnetic vector potential, for visualization purposes the data was smoothed.} 
\label{fig:weibel}}
\end{figure}

Interaction between the \mpon{incoming and the shock-reflected ions triggers} the Weibel instability. It strongly amplifies the magnetic field and changes its structure at the shock foot and at the ramp. Moreover, the Weibel filaments may become tearing-mode unstable and decay through magnetic reconnection \citep{Matsumoto2015,Bohdan2020a}. This creates magnetic vortices, which \mpon{may collate} into larger structures. 

The top panel in Figure \ref{fig:weibel} compares the evolution of the energy density of the Weibel-generated magnetic field \KF{for all runs. To calculate this quantity, we consider the magnetic field in the region where the Weibel instability operates: $-5\SkinI<x-x_{sh}<15\SkinI$. Then, the large-scale gradients are removed to exclude the effect of compression, and the Fourier power spectrum is computed. The energy density is obtained directly from the spectrum, and it} is normalized to the energy density of the initial magnetic field $U_{B_0}=|B_0|^2/(2\mu_0)$. \mpon{To be noted are the cycle-to-cycle variations in the} quasi-periodic behavior caused by the shock reformation. On average there is no significant difference in the energy density \mpon{with and without upstream turbulence, and likewise} in the magnetic reconnection activity, since it depends on the strength of the Weibel instability. Indeed, the bottom panel in Figure \ref{fig:weibel} \mpon{indicates a similar number of magnetic vortices for \KF{all} runs, traced here as} \kf{local maxima in $z$-component of the magnetic vector potential.}

\begin{figure*}[ht!]
\includegraphics[width=\textwidth]{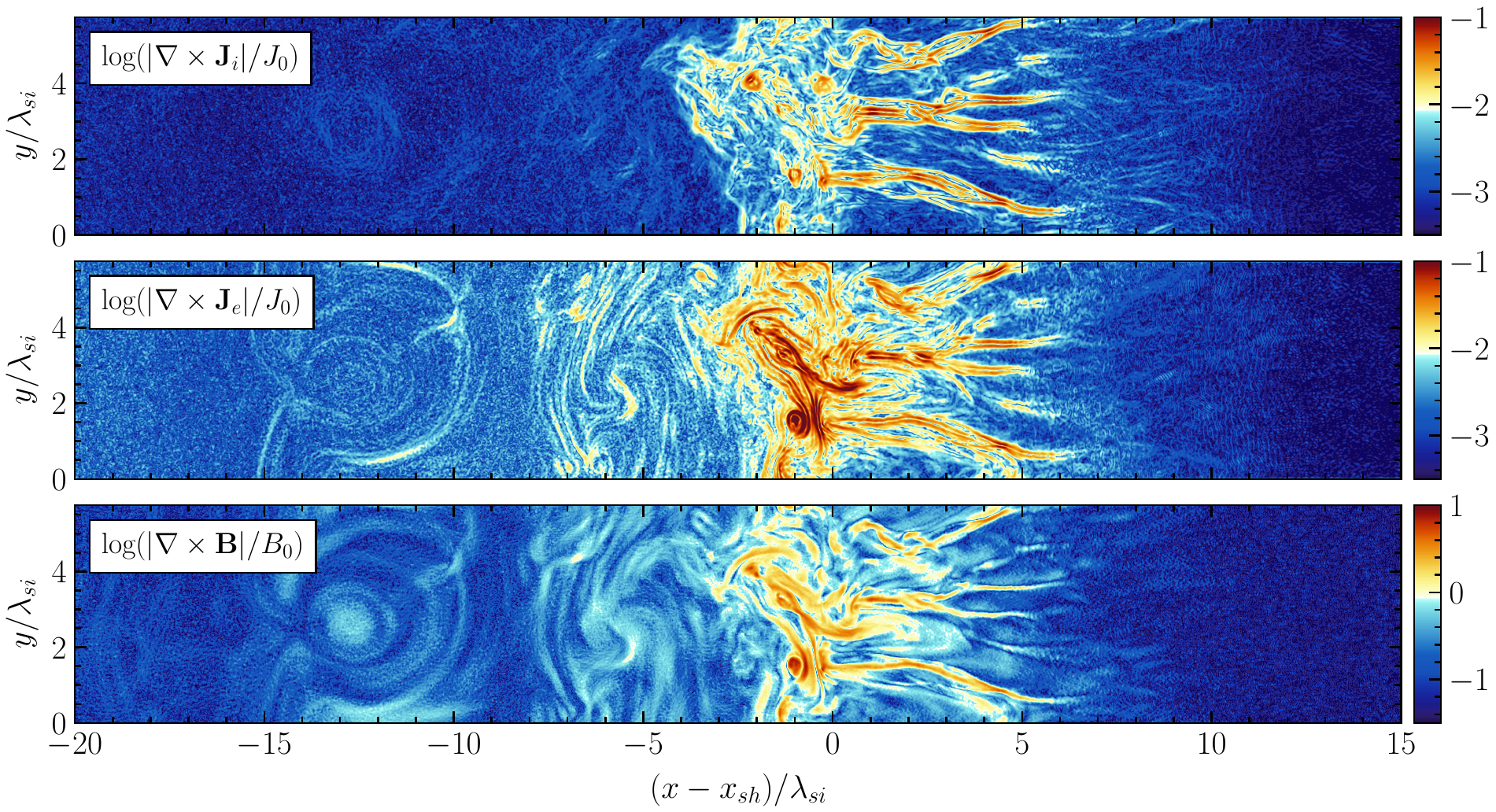}
\caption{Maps of the \kf{magnitude of the curl of the electron current density (\textit{top} panel), the ion current density (\textit{middle} panel), and the magnetic field (\textit{bottom} panel), for run T2 in the middle of the third reformation cycle ($\sim 8.5 \LarmorT$). The current densities are normalised by the factor $J_0=n_0 q_e c$.} 
\label{fig:curl}}
\end{figure*}

\subsection{Vorticity}\label{sec:vorticity}

In the MHD regime, the postshock magnetic field can be amplified via the turbulent dynamo \citep{Giacalone2007,Fraschetti2013,Xu2016,Hu2022}. In this \mpon{process}, the interaction of a shock with pre-existing density inhomogeneities causes corrugations of the shock front in the form of ripples, which generate strong rotation and vorticity of the downstream fluid. The frozen-in magnetic field lines are bent and distorted, and that leads to amplification of the field. For this reason, we investigate vorticity\kf{, specifically the curl of the particle current density,} on the kinetic scales.

Figure \ref{fig:curl} shows maps of \kf{the magnitude of the curl of the current density} and the magnetic field, calculated for the run with turbulence (T2), at the stage of the reformation cycle when the Weibel filaments are well developed \KF{($t \approx 8.5 \LarmorT$)}. The subscripts ``$e$'' and ``$i$'' refer to electrons and the ions, respectively. The pre-existing density fluctuations do not carry significant vorticity, hence in the upstream the curl of the currents and the magnetic field is weak. At the shock foot, the vorticity is dominated by the Weibel filaments. \KF{The dense filaments found between $x-x_{sh} = (0-5)\SkinI$ are composed of plasma moving towards the shock. They are surrounded by the stream of the shock-reflected particles, inducing shear flows that cause strong curl of both $\textbf{J}_i$ and $\textbf{J}_e$ at the filament contours.} \KF{The current density inside the filaments produces the strong curl of the magnetic field, which is \mponn{seen} in the bottom panel of Figure \ref{fig:curl} in that region.} The Weibel filaments are unstable and dissipate via magnetic reconnection. This is visible as circular structures both in ion and electron current, as well as local maxima of the curl of the magnetic field. The magnetic vortices converge into larger ones and travel through the shock further downstream. A particularly large magnetic vortex lies between $-15\SkinI<x-x_{sh}<-11\SkinI$. The corresponding current comes only from electrons since the dynamic scale of the shock-transmitted ions is much larger than the scale of the magnetic field structure. 

The vorticity strongly varies \mpon{between one reformation cycle and the next}. \kf{We observed a large vortex \mpon{in the downstream region} only for run T2 during the third reformation cycle. We find no correlation between the \mpon{presence of the large vortex and the} magnetic reconnection activity, which indicates serendipity of this event.} On average we do not observe a significant difference between the vorticity level in simulations with and without density fluctuations in the upstream plasma. The vorticity \mpon{seems to arise} only from magnetic reconnection at the Weibel filaments. This may be different if the scale of the density clumps is larger than the shock width \mpon{or the scale of rippling instabilities. Study of the latter would require a correspondingly wide simulation box.}

\subsection{Obliquity angle}\label{sec:obliquity}

\begin{figure}[t!]
\centering 
\includegraphics[width=\columnwidth]{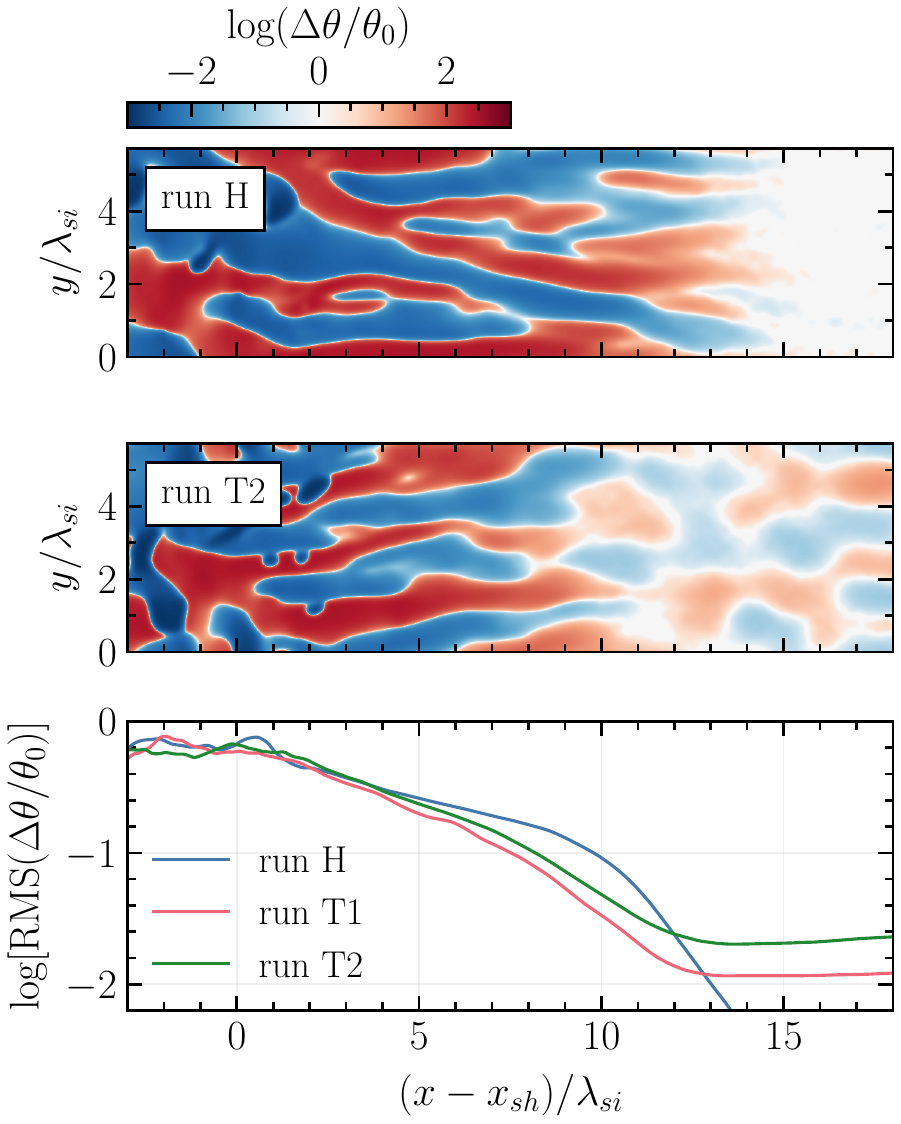}
\caption{\KF{Variations of the obliquity angle in the region extending from the upstream to the overshoot. The upper and middle panels show maps of the difference between the local obliquity angle and the inclination angle of the external magnetic field for the runs H and T2, respectively, and the time $t\approx8.8\LarmorT$. The maps are presented in logarithmic and sign-preserving scale: $\text{sgn}(\Delta \theta)\cdot[2.5+\log \{\text{max}(10^{-2.5},|\Delta \theta|/\theta_0)\}]$. The lower panel shows the RMS of the difference between the local obliquity angle and the inclination angle of the external magnetic field for each run. The profiles are obtained from $\Delta\theta$ maps by calculating the RMS over the $y$-direction for each time step in the latest reformation cycle. Then the quantity is time-averaged and normalized by the factor $\theta_0$.}
\label{fig:theta}}
\end{figure}

Variations of the magnetic field ahead of the shock front may lead to local changes of the shock obliquity angle. This means that reflection conditions of particles are modified. In this section, we examine whether the \mpon{variance of the obliquity angle in the foot region depends on the presence of} pre-existing density fluctuations.

Figure \ref{fig:theta} shows 2D maps \kf{of the difference between the local obliquity angle and the inclination angle of the external magnetic field: \KF{$\Delta \theta = \theta-\theta_{0}$,}} for run H (the middle panel) and T2 (the bottom panel), \KF{at the time $t\approx8.8\LarmorT$}.
Additionally, the root mean square of this quantity calculated over the $y$-direction is included. The RMS value for run H ahead of the shock foot, $x \geq 15\SkinI$, is \mponn{very much smaller than that in} \KF{the runs with turbulence it approximately equals: $1.1^\circ$ (T1) and $2.3^\circ$ (T2).} Even so, the variation of the obliquity angle becomes similar for \mponn{all} runs in the regions closer to the shock front: at the foot, at the ramp and at the overshoot. The structure of the magnetic field is strongly modified there by the Weibel instability. The similar values of the RMS for all runs indicates that the instability is not affected by the upstream turbulence (see Section \ref{sec:weibel}). 

\subsection{Transmission of fluctuations}\label{sec:fluc_trans}

\begin{figure}[t!]
\includegraphics[width=\columnwidth]{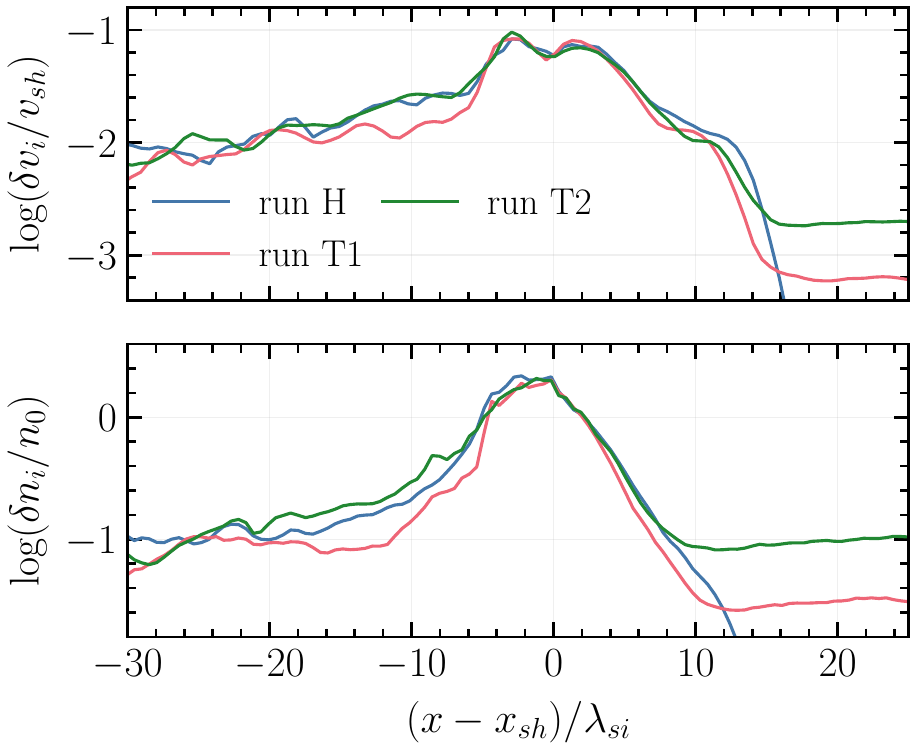}
\caption{\KF{Level of the speed and density fluctuations for all simulations in the region extending from upstream to downstream of the shock. The upper panel shows the RMS of the ion bulk speed fluctuations normalised to the shock speed, and the lower panel shows the RMS of the ion density fluctuations normalised to the initial upstream density. The profiles are obtained from the velocity and the density maps by calculating the RMS over the $y$-direction and combining two neighboring cells in the $x$-direction. The profiles are time-averaged over the latest reformation cycle.}
\label{fig:turb_trans}}
\end{figure}

\mpon{Having discussed the influence of the upstream density fluctuations on the shock waves, we now explore} how the turbulence \mpon{is} transmitted across the shock, focusing on the amplitude of the velocity \mpon{and density} fluctuations.

The upper panel in Figure \ref{fig:turb_trans} shows the \mpon{amplitude} of fluctuations of the ion bulk speed, \KF{at scales of roughly $3\SkinE \times 3\SkinE$. The RMS of the fluctuations, $\delta v_i$, was calculated in regions of the box} \mpon{width,  $\Delta y\approx 60\SkinE$, \KF{and combining two neighbouring cells in the $x$-direction.}} The pre-existing turbulence \KF{in run T2} carries \KF{velocity fluctuations of roughly 0.2\% of the shock speed.} \KF{For run T1 the amplitude is approximately smaller by factor 3}. Nevertheless, \KF{in all simulations, the amplitude of the fluctuations at the shock ramp and further downstream of the shock reaches roughly 10\% and 1\% of the shock speed, respectively.}

The bottom panel in Figure \ref{fig:turb_trans} presents the  \mpon{amplitude} of density fluctuations for all simulations. \KF{The amplitude $\delta n_i$ is calculated in the same way as for $\delta v_i$.} In the upstream region, $x-x_{sh}>12\SkinI$, the initial fluctuation level carried by the pre-existing turbulence \KF{are equal to $\delta n_i/n_0\simeq 3.5 \%$ and $\delta n_i/n_0\simeq 10 \%$ for runs T1 and T2, respectively. This corresponds to the turbulence amplitudes from Table \ref{tab:sim_params}.}  \KF{Furthermore, there is no significant spatial variation of the turbulence level along the $x$-axis ahead of the shock foot. This sustains that the statistical properties of the fluctuations are stable.} \KF{For runs T1 and T2 the amplitude of the density fluctuations increases gradually from  $x-x_{sh}\simeq10\SkinI$ and $x-x_{sh}\simeq9\SkinI$, respectively,} whereas for the homogeneous setup, \KF{it happens at the beginning of the shock foot, $x-x_{sh}\simeq12\SkinI$. This amplitude growth is caused by the formation of the Weibel filaments, which at some distance surpass the pre-existing turbulence.} For all cases, the fluctuation amplitude reaches $\delta n_i/n_0 \approx 250\%$ at the \KF{overshoot and \mponn{then decreases to} $\delta n_i/n_0 \approx 10\%$ further in the downstream.}

Recent 2D hybrid simulations of a shock wave propagating in a high-$\beta$ turbulent medium show significant amplification of the density and the velocity fluctuations \citep{Nakanotani2022} \mpon{for density structures of a few tens of ion skin lengths \KF{and amplitudes of unity of the upstream density}, whereas we see little effect of upstream density variations that are about ten times smaller. \mponn{We might} expect analogous physical features at both scales,} \mpon{but with a smaller size any forcing is applied for a correspondingly shorter time, hence the smaller impact}. In our simulations, the magnitude of fluctuations \KF{in the shock transition and in the downstream regions is independent of the strength of the pre-existing turbulence. This suggests that the level of the fluctuations is dominated by processes occurring in the shock transition (e.g., the Weibel instability, compression)}.

\subsection{Downstream electron temperature}\label{sec:temperature}

\begin{figure}[t!]
\includegraphics[width=\columnwidth]{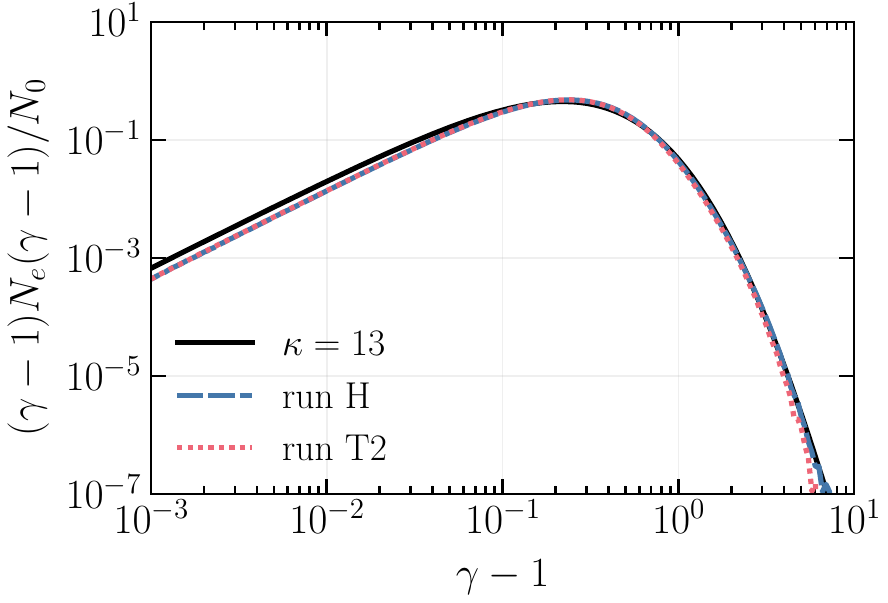}
\caption{The electron energy spectrum in the downstream region for runs \kf{H (the blue dashed line) and T2 (the red dotted line)}. The black solid line denotes a fitted \kff{kappa} distribution.
\label{fig:temp}}
\end{figure}

Figure \ref{fig:temp} \mponn{compares} the electron energy spectra in the far downstream region, $x-x_{sh}=-(32\text{ -- }64)\SkinI$, for the runs \mponn{with homogeneous plasma (H) and} with $\delta n/n =10\%$ (T2). The \mpon{energy distribution is calculated for the late stage of the simulation and} in the local plasma rest frame. The electron spectra \kff{are represented by a kappa distribution with $\kappa=13$, denoted by the black solid line, demonstrating}
\mpon{a weak high-energy tail that is virtually the same in} the simulations with and without upstream turbulence. In general, the shape of the downstream electron spectra is a consequence of particle interactions \mpon{such as \mponn{shock-surfing acceleration} and magnetic reconnection. The similarity of the spectra reflects the absence of significant modifications of the main instabilities at the shock}.
 
The temperature of electrons is calculated from the fitted \mpon{thermal distribution and is} presented in Table \ref{tab:sim_params}. Within the uncertainties, \mpon{and averaging over the spatial variation of the downstream temperature, we find the same} temperature in all runs. The obtained electron temperatures are in agreement with the value $0.183\pm0.004$ \mpon{reported} for run B2 in \citet{Bohdan2019b}. The electron heating model of \citet{Bohdan2020b} includes a combination of different processes operating \ab{in the shock transition}, \mpon{modification of which by small-scale density fluctuations was not observed, suggesting that there is no impact on} electron heating.

\KF{We do not discuss the temperature of ions because the duration of our simulations is too short for these particles to reach 
equilibrium. \kfff{However, findings from 1D PIC simulations suggest that spectra of thermal and suprathermal ion populations in the downstream region of quasi-parallel shocks can be represented by a kappa distribution \citep{Arbutina2021}.}}

\section{Summary and discussion} \label{sec:summary}
Understanding the physics of a plasma shock wave propagating in an inhomogeneous medium is an important challenge in astrophysics and space physics. We present results of kinetic simulations of high-Mach-number shocks with pre-existing density fluctuations, which model conditions at SNRs. Our main goal was to investigate the influence of the turbulence on electron-scale phenomena: the Buneman and the Weibel instabilities, electron acceleration, and their heating. The second aim was to examine the global shock structure, including the properties of the magnetic field. Furthermore, we analysed how the upstream fluctuations are transmitted across the shock. Our results can be summarised as follows:
\begin{enumerate}
    \item Our novel simulation technique establishes a framework for studying shocks propagating in inhomogeneous media. It may be used to model various physical environments \mpon{and does not bear significant} computational costs.
    \item \kff{The achievable level of pre-existing upstream turbulence on kinetic scales is limited by particle heating. Our empirical test indicate that to maintain $M_S\gtrsim30$ the maximum amplitude of density fluctuations should be of the order of $\delta n/n\sim10\%$ (on the scale of a quarter of the ion skin length). The limit should be even lower for shock speeds closer to those observed in SNRs.}
    \item We performed two simulations of a perpendicular high-Mach-number shock with different \mpon{RMS amplitude} of pre-existing density fluctuations, $\delta n/n=3.5\%$ and $\delta n/n=10\%$. The amplitudes were chosen \mpon{to avoid excessive heating and in concordance with in-situ measurements}. We examined the impact of the turbulence on the shock physics and the electron dynamics by direct comparison to \mpon{a simulation with} homogeneous upstream \mpon{medium}.
    \item \mpon{Density fluctuations up to a few $\SkinI$} do not change the global structure of a perpendicular shock wave, as well as its reformation period and propagation speed. No significant difference was found in the amplification of the magnetic field, \mpon{on account of weak} turbulence-driven vorticity on kinetic scales.
    \item Fourier analysis of the Buneman waves in the shock foot shows that their properties remain unchanged under interaction with the density inhomogeneities. Likewise for the Weibel modes and magnetic reconnection at them. The magnetic field fluctuations carried by the pre-existing turbulence locally modify the magnetic field obliquity at the shock foot. At the ramp the topology of the magnetic field is dominated by the Weibel instability.
    \item The velocity and the density fluctuations \mpon{in the downstream region reach a comparable} amplitude for all \mpon{levels of pre-existing density fluctuations that we studied}.
    \item Energy spectra of downstream electrons \mpon{are thermal with a weak tail\kff{, but the electron pre-acceleration is not sufficient for injection into DSA.} The temperature and non-thermal fraction} are similar for all simulations, and electron acceleration and heating are unmodified.
\end{enumerate}

Taken together, these findings suggest that pre-existing density fluctuations \mpon{of a few $\SkinI$ and with} realistic intensity do not considerably affect the physics of high-Mach-number shocks. \kff{In particular, we did not observe any changes in the properties of plasma instabilities or in the efficiency of electron acceleration.} \mpon{This may be different for turbulence on larger scales, as hybrid simulations suggest  \citep{Trotta2021,Nakanotani2022,Trotta2023}. 
There turbulent structures introduced by shock corrugation enhance particle acceleration efficiency, as well as amplify the magnetic field via the turbulent dynamo. Analogous studies in kinetic regime require using a wider simulation box and thus more computational resources.}

The density fluctuations are also too weak to impact the ion reflection dynamics, which also could modify the instability conditions. On the other hand, the density structures are comparable in size to the Weibel filaments. Although, they have a much smaller amplitude, so they do not modify the Weibel instability. \mpon{At lower Alfv{\'e}nic Mach numbers the Weibel modes should be weaker, and correspondingly the impact of pre-existing turbulence larger.}

\kf{Further studies should explore the effect of the Alfv{\'e}nic turbulence, where the magnetic field fluctuations dominate.}



In contrast to perpendicular shocks, \mpon{from which electrons cannot escape to the upstream region, at oblique shocks the particles form extended foreshock, where they drive instabilities \citep{Bohdan2022,Morris2022}. This adds room for} particle-turbulence interactions, and hence changes of the reflection conditions and the electron dynamics, as well as modifications of the instabilities. \kff{Moreover, 1D3V \citep{Xu2020,Kumar2021} and 2D3V \citep{Morris2023} simulations show that a significant non-thermal electron population can be formed \kff{downstream} of such shocks \kff{that} is well described by a power-law. Therefore, at oblique shocks it is possible to directly investigate \kff{the} influence of pre-existing turbulence on energetic particles.}


\kff{Full 3D simulations are currently \kff{overwhelmingly} challenging, \kff{and} so they must be run with lower resolution and smaller boxes and can only reach \kff{the very early} stages in the evolution of the system \citep{Matsumoto2017}, in contrast to 2D3V simulations \citep{Bohdan2017}. Nevertheless, including all dimensions may influence the interplay and nonlinear development of the instabilities driven at the shock, \kff{and} cross-field diffusion can be adequately captured, as recently argued by \citet{Orusa2023}.}

\vspace{0.5cm}
\noindent
K.F. and M.P. acknowledge support by DFG through  grants PO 1508/10-1 and PO 1508/11-1. This research was supported by the International Space Science Institute (ISSI) in Bern, through ISSI International Team project \#520
\textit{Energy Partition across collisionless shocks}. A.B. was supported by the German Research Foundation (DFG) as part of the Excellence Strategy of the federal and state governments - EXC 2094 - 390783311. G.~T.~P. acknowledge support by Narodowe Centrum Nauki through research project No. 2019/33/B/ST9/02569. The numerical simulations were conducted on resources provided by the North-German Supercomputing Alliance (HLRN) underproject bbp00057. We gratefully acknowledge Polish high-performance computing infrastructure PLGrid (HPC Centers: ACK Cyfronet AGH) for providing computer facilities and support within computational grant no. PLG/2022/015967.

\newpage
\appendix

\begin{figure*}[ht!]
 \centering 
 \includegraphics[width=0.43\linewidth]{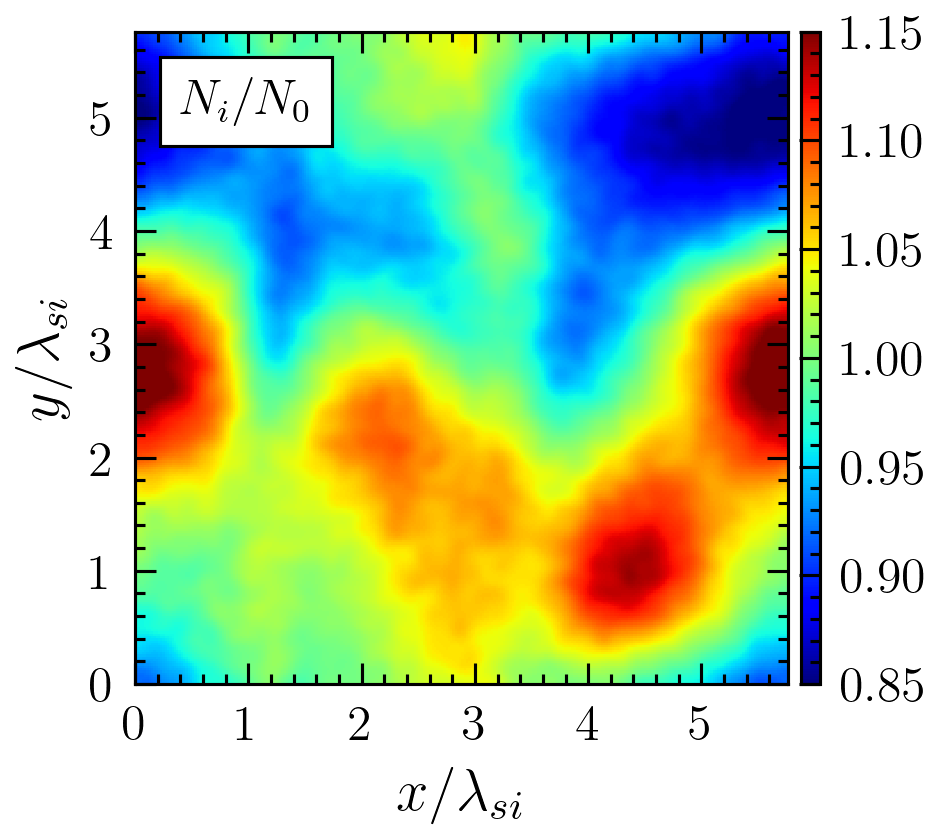} 
 \hfil 
 \includegraphics[width=0.5\linewidth]{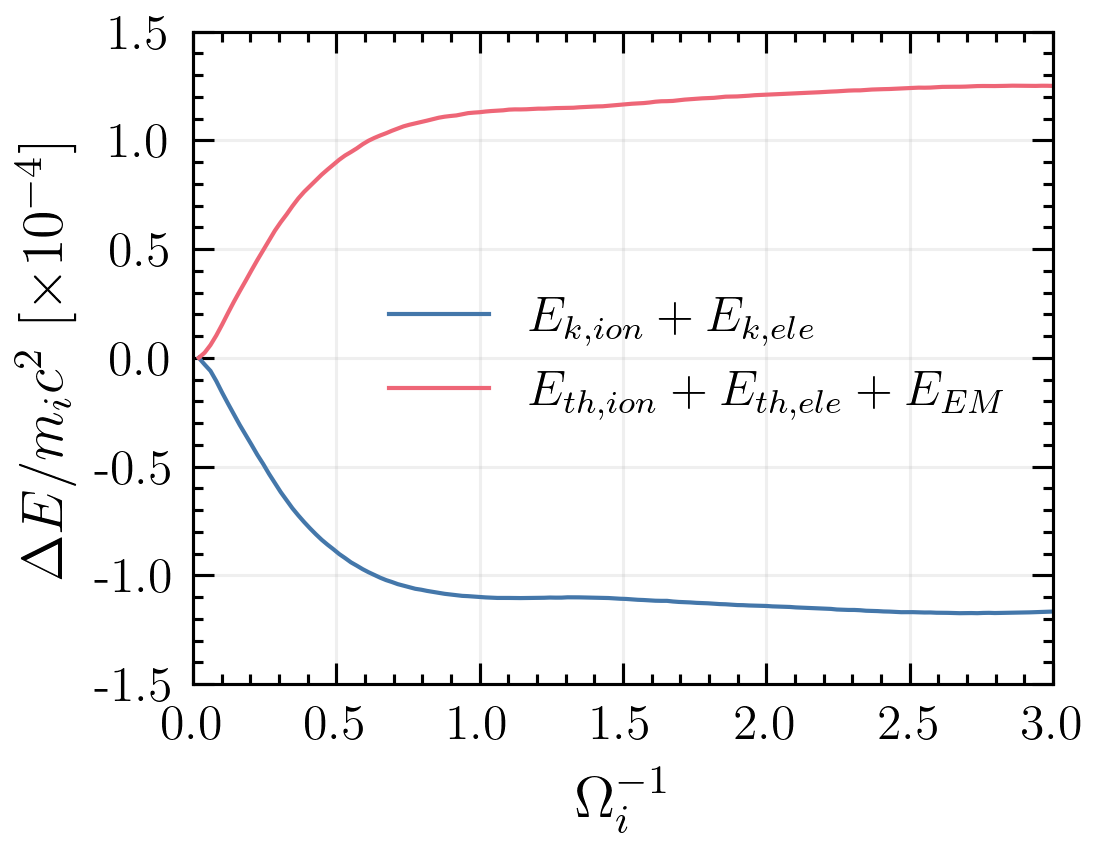} 
\caption{Left: Normalized ion number density map showing well-developed structure of the compressible turbulence, after $1\LarmorT$. Right: Time evolution of the mean kinetic energy, as well as sum of the thermal energy and the electromagnetic field energy. The initial values are subtracted, the energies are normalized to the ion rest energy.
\label{fig:slab}}
\end{figure*}

\section{Generation of compressive turbulence}
In this work, we consider decaying and compressive turbulence, which is established by initially imposing a local disturbance of the bulk velocity. The velocity of the plasma species at the arbitrary location $\mathbf{v(r)}$, contains the bulk component, its disturbance, and the thermal fluctuations: $\mathbf{v(r)}=\mathbf{v_0}+\delta \mathbf{v_0(r)}+\mathbf{v_{th}}$. The disturbance is in the form of the superposition of wave-like modes, with both the transversal and the longitudinal configuration. This is similar to techniques from \citet{Giacalone1994} and \citet{Groselj2019}, but we only impose the velocity disturbances without any magnetic field associated with them. The form of the transversal disturbance is given by the following equation:

\begin{equation}
    \delta \mathbf{v_0(r)} = \sum_k A_k\cos(kx'+\beta_k)\hat{y}'+A\sin(kx'+\beta_k)\hat{z}',
\end{equation}
and for the longitudinal one:

\begin{equation}
    \delta \mathbf{v_0(r)} = \sum_k A_k\cos(kx'+\beta_k)\hat{x}'.
\end{equation}
The relation between the primed system and the unprimed is given by the rotation matrix

\begin{equation}
\mathbf{r}' = 
\begin{pmatrix}
\cos\phi & \sin\phi & 0 \\
-\sin\phi & \cos\phi & 0 \\
0 & 0 & 1
\end{pmatrix}
\mathbf{r}.
\end{equation}

For each randomly chosen wavenumber $k$ we chose a random orientation with respect to the $x$ axis: $0<\phi_k<2\pi$, and a random phase $0<\beta_k<2\pi$. Letter $A_k$ denotes the amplitude of each wave. Periodic boundaries require that the wavevector components have to fulfill the following identities: $k_xL=n\pi$ and $k_yL=m\pi$, where $L$ is the box size and $n,m$ are random integers. We note that the initial spectrum strongly depends on combinations of the $n,m$ pairs, but it changes as the system evolve. As we mention above, we do not impose any magnetic field disturbance, but since initially the magnetic field has a constant value, we must input the correct motional electric field $\mathbf{E}_0=-\mathbf{v} \times \mathbf{B}_0$.

In our simulations, we generate 50 waves (longitudinal or transversal), with randomly chosen $n,m$ such that the wavevector component is in the range $\frac{2\pi}{L}<k_x<6\cdot\frac{2\pi}{L}$ (same for $k_y$). These pairs of integer numbers give us the orientation of each wave $\phi_k$. Phase of the wave $\beta_k$ is chosen randomly, furthermore, we assume that all amplitudes are equal: $A_k=A=\text{const}$. This setup efficiently generates compressive turbulence, which intensity can be adjusted by the parameter $A$. The well-developed density structure of turbulence with the intensity $\delta n/n \approx 10\%$ is presented in Figure \ref{fig:slab} (the left panel). The decay time of the fluctuations exceeds $2\LarmorT$, so there are sufficiently long-lived to be inserted into a shock simulation. The right panel in this figure shows the time evolution of the particle kinetic energy $E_k$, and the sum of the particle thermal energy and the energy of the electromagnetic field $E_{th}+E_{EM}$. The energies are averaged over the simulation box, their initial values are subtracted, and they are normalized to the ion rest energy. Our kinetic simulations cover the dissipation range of turbulence, so the energy of the initial fluctuations is partially transferred into plasma heat.  The sonic Mach number of a shock is defined by the end state of the particle temperature, hence the heating constraints the maximum achievable turbulence level for a given Mach number.

\section{Matching turbulent plasmas}

\begin{figure*}[t!]
 \centering 
 \includegraphics[width=0.55\linewidth]{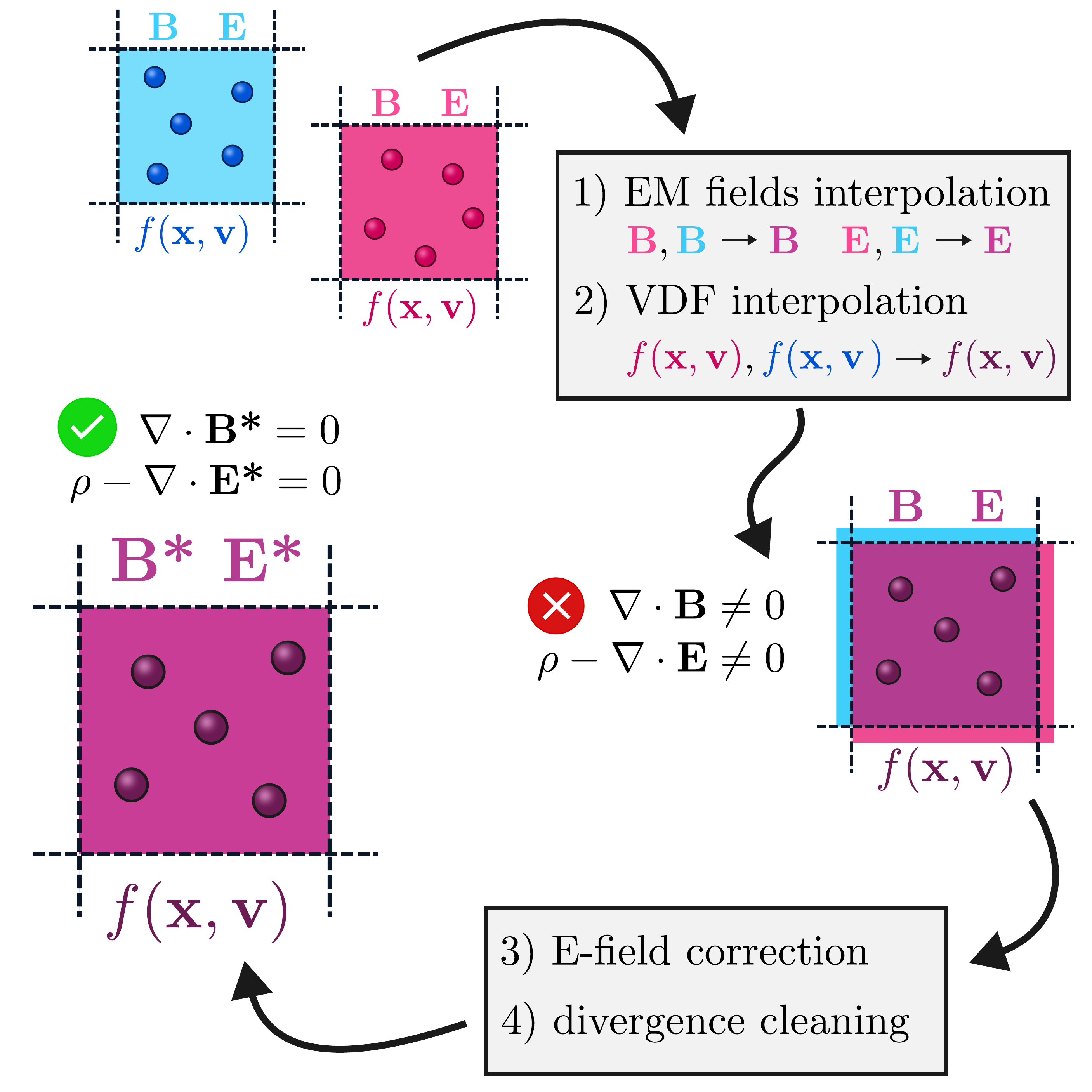}
 \hfill
 \includegraphics[width=0.4\linewidth]{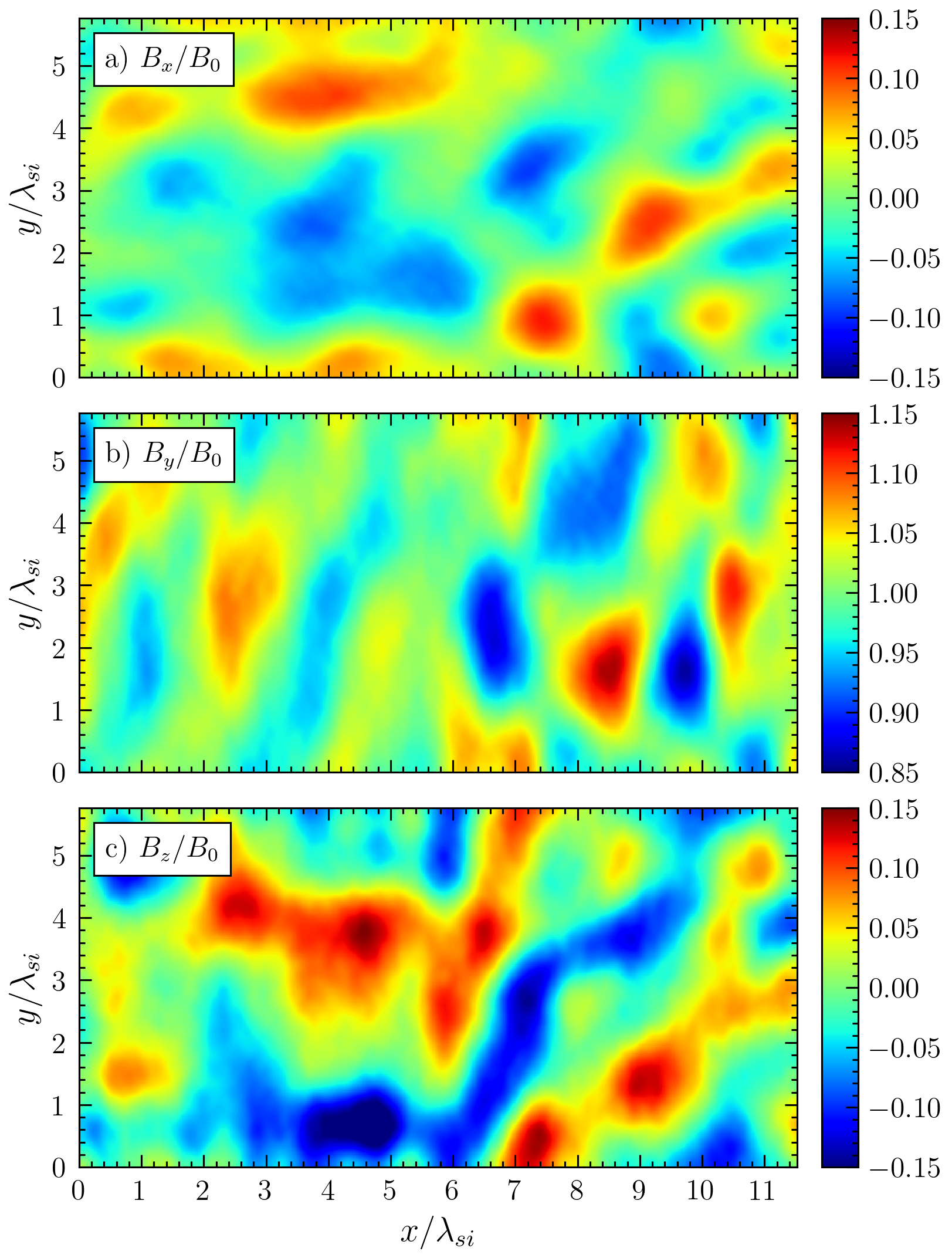}
\caption{\KF{Left: The matching procedure presented for one grid cell.} Right: Maps of the magnetic field after around half an ion Larmor time after merging. The interpolation zone spans from $x=5.3\SkinI$ to $x=5.8\SkinI$ (middle of the panel).
\label{fig:matching}}
\end{figure*}

Injection of a pre-fabricated turbulent plasma slab into a shock simulation requires matching of the plasma at the interface. Otherwise, any mismatch in the electromagnetic field and the current density may drive artificial transients. Here, we discuss the matching of two arbitrary plasma slabs, called the left and the right, pre-fabricated in simulations with periodic boundaries. We chose a region where the electromagnetic fields and the current densities are interpolated: the interpolation zone. The values of the fields in each grid cell of the zone are replaced by the interpolated ones, considering the magnetic field:
\begin{equation}
    B(i,j) = w(i) \cdot B_L(i,j)+[1-w(i)] \cdot B_R(i,j),
\end{equation}
where the subscripts ``$L$'' and ``$R$'' denote the left and the right slab, $i,j$ are the cell numbers in $x$ and $y$-direction respectively, and $B$ is the interpolated value of the magnetic field. The weights $w(i)$ depend only on the $i$-th location on the simulation grid, and they are imposed in the form of a linear function. The same procedure applies to the electric field. Simple calculations show that the field interpolation does not contribute to large factors in both Amp\`{e}re's and Faraday's law.

Despite the fields, the current densities also need to be matched. In PIC codes the current is deposited by particles, thus a smooth transition in the velocity distribution function of plasma species is required. The distribution of each component of the particle velocity vector is assumed as the normal distribution. Per each cell of the slabs in the interpolation zone, the first three moments of the distribution are calculated, and multiplied by proper weights, which gives us the interpolated moments. Afterwards, all particles in the zone are deleted and a new distribution is reestablished from the interpolated moments. For the reason that the number of particles per cell is not large, $n_{ppc}=20$, to reduce statistical noise $3\times3$ cells are used to calculate the moments.

Demand that the first moment of the distribution function (the number of particles) must be an integer gives a rounding error, which leads to an artificial charge noise: $\delta \rho =  \rho - \nabla \cdot \mathbf{E}$. This must be corrected to satisfy Gauss law, because it is not explicitly solved in the code. We add an additional component to the electric field, $\mathbf{E}^*=\mathbf{E}+\delta\mathbf{E}$, such that the corrected field satisfies $\rho - \nabla \cdot \mathbf{E}^*=0$. The additional field $\delta\mathbf{E}$ is obtained from the superposition principle.

Another correction must be done for the magnetic field. The weights $w(i)$ depend on the $i$-th location on the computational grid, which produces a nonzero magnetic field divergence: $\nabla \cdot\mathbf{B} \neq 0$. To clean the divergence the projection method is used \citep[see Section 3.3 in][]{Zhang2016}. We briefly present here how the method works. The magnetic field can be written as the sum of a curl and a gradient component:
\begin{equation}
    \mathbf{B} = \mathbf{\nabla} \times \mathbf{A} + \nabla \phi,
\end{equation}
where $\mathbf{A}$ is a vector field and $\phi$ is a potential. After taking the divergence of both sides we obtain the Poisson equation:
\begin{equation} \label{eq:Poisson}
    \nabla^2 \phi = \nabla \cdot \mathbf{B}.
\end{equation}
If we calculate the potential and extract its divergence from the initial magnetic field, we get the corrected magnetic field:
\begin{equation}
    \mathbf{B^*} = \mathbf{B} - \nabla \phi.
\end{equation}
We solve Equation \ref{eq:Poisson} by the Full Multigrid Algorithm described in \citet[p. 868-874]{NR1992}, and implemented in FORTRAN 90 in \citet[p. 1334-1338]{NR1996}.

\KF{The matching procedure is illustrated in the left panel of Figure \ref{fig:matching}.} We tested our method for the interpolation zone of 100-cell thickness. The right panel in Figure \ref{fig:matching} \KF{highlights} that any electromagnetic transient does not form.

\bibliography{references}{}
\bibliographystyle{aasjournal}

\end{document}